\documentclass[prl,aps,10pt,twocolumn]{revtex4-1}
\usepackage[dvipsnames]{xcolor}		
\usepackage{amsmath}    
\usepackage{graphicx}   
\usepackage{verbatim}   
\usepackage{color}      
\usepackage{subfigure}  
\usepackage[hidelinks]{hyperref}   
\usepackage[locale=US]{siunitx}
\usepackage{booktabs}
\usepackage{pgfplots}
\usetikzlibrary[pgfplots.groupplots] 
\usepgfplotslibrary{external} 
\usepgfplotslibrary{fillbetween} 
\usetikzlibrary{automata,positioning,calc,backgrounds}
\usepackage{braket}
\usepackage{placeins}
\usetikzlibrary{shadows}
\tikzset{
    text shadow/.code args={[#1]#2at#3(#4)#5}{
        \pgfkeysalso{/tikz/.cd,#1}%
        \foreach \angle in {0,5,...,359}{
                \node[#1,text=blue] at ([shift={(\angle:.20pt)}] #4){#5};
        }
    }
}
\def\k{0.55191496}

\tikzset{
    sphere color/.store in=\spherecolor,
    sphere scale/.store in=\spherescale,
    sphere color=blue,
    sphere scale=1,
    sphere/.style={
        ultra thick,
        line join=round,
        draw=#1!75!black,
        ball color=#1,
    },
    sphere inside/.style={
        shading angle=180,
        sphere=#1!25!gray!75!black
    }
}

\newenvironment{sphere}[1][]
    {
        \begin{scope}[x=(0:1cm), y=(90:1cm), z=(260:0.25cm), #1]
            \path [sphere inside=\spherecolor, scale=\spherescale] 
            circle [radius=1];
    }
    {
        \path let \n1={cos 10}, \n2={sin 10} in [sphere=\spherecolor, scale=\spherescale, even odd rule, opacity=0.5]
        circle [radius=1] 
        [x={(\n1, \n2^2, \n2*\n1)},
         y={(0, \n1, \n2)}, 
         z={(-\n2, -\n1*\n2, \n1^2)}] (0,1,0) 
                .. controls ++( 0, 0,\k) and ++(0,\k, 0) .. (0, 0, 1)
                .. controls ++(\k, 0, 0) and ++(0, 0,\k) .. (1, 0, 0) 
                .. controls ++(0, \k, 0) and ++(\k,0, 0) .. (0, 1, 0);
        \end{scope}
    }

\begin{document}
%
\title{Investigation of room temperature multispin-assisted bulk diamond \textsuperscript{13}C~hyperpolarization at low magnetic fields}
\author{Ralf Wunderlich\textsuperscript{1}, Jonas Kohlrautz\textsuperscript{1}, Bernd Abel\textsuperscript{2}, J\"urgen Haase\textsuperscript{1}, Jan Meijer\textsuperscript{1}*}
\affiliation{\textsuperscript{1}\textit{Faculty of Physics and Earth Sciences, Felix Bloch Institute for Solid State Physics, Leipzig University, Linnéstrasse 5, 04103 Leipzig, Germany}} 
\affiliation{\textsuperscript{2}\textit{Leibniz-Institute of Surface Engineering (IOM), Permoserstrasse 15, 04318 Leipzig, Germany}} 
\affiliation{*\textbf{Corresponding author}: ralf.wunderlich@uni-leipzig.de} 
\date{\today}
\begin{abstract}
In this work we investigated the time behavior of the polarization of bulk $^{13}$C~nuclei in diamond above the thermal equilibrium. This nonthermal nuclear hyperpolarization is achieved by cross relaxation between two nitrogen related paramagnetic defect species in diamond in combination with optical pumping.
The decay of the hyperpolarization at four different magnetic fields is measured.
Furthermore, we use the comparison with conventional nuclear resonance measurements to identify the involved distances of the nuclear spin with respect to the defects and therefore the coupling strengths.
Also, a careful look at the linewidth of the signal give valuable information to piece together the puzzle of the hyperpolarization mechanism.
\end{abstract}
\maketitle
\section{Introduction}
Any spin resonance technique, including nuclear magnetic resonance (NMR), is based on the occupation difference of the energy levels associated with different magnetic quantum numbers at a given magnetic field. Unfortunately, this is given by the Boltzmann distribution and leads to a tiny occupation difference at room temperature.
In recent years, more and more experimental as well as theoretical contributions are published in the literature dealing with nuclear hyperpolarization utilizing the negatively charged nitrogen vacancy (NV) center in diamond \cite{Jacques2009,Wang2013,Fischer2013a,King2015,Alvarez2015,
Ivady2015,Yang2016,Scheuer2016,FernandezAcebal2017}.
Recently, a nuclear hyperpolarization method without the need of microwave application was presented \cite{Wunderlich2017,Pagliero2018}. There, it is shown that cross relaxation (CR) between NV centers and subsitutional nitrogen (P1) centers leads to a $^{13}$C~hyperpolarization in several narrow magnetic field regions in the range of \SIrange{48.5}{53.5}{\milli\tesla}. An additional advantage of this technique is the  applicability of type I diamonds without the need of ultra pure and expensive samples. 
Here, we investigate the time dynamics of this method. Furthermore, a comparison of hyperpolarized signals with conventional measurements in the thermodynamic equilibrium (TE) will be used to identify the polarised regions with respect to the paramagnetic defects.
\section{Results}
\subsection{Time dependence of nuclear hyperpolarization}
The experimental setup for the hyperpolarization measurements is described in detail in Ref.~\citenum{Wunderlich2017}.
We used a single crystal diamond sample with an estimated nitrogen content of \SI{200}{ppm}, primarily present as P1 centers. The NV density is estimated to be maximal in the range of several ppm.
The crystallographic $[111]$ direction of the diamond sample was set parallel to the external magnetic field.
All nuclear free induction decay (FID) signals were recorded at \SI{7.05}{\tesla} (\SI{300}{\mega\hertz} proton Larmor frequency), if not described differently. The advantage of using bulk NMR techniques in comparison to optically detected magnetic resonance (ODMR) is that this technique is sensitive to all nuclear spins, even if they are far away from any ODMR active center.\par
In the following, we present the characteristic build up time as well as the decay time at four different magnetic fields of the nuclear hyperpolarization.
\begin{figure}[b!t]
\centering
\tikzsetnextfilename{pumpingtime}
\begin{tikzpicture}[background rectangle/.style={fill=white, opacity=1.0}, show background rectangle]
		\begin{axis}[name=plot1,
		height=5.0cm, width=0.8\linewidth,
		xlabel={illumination time (\si{\second})},
		ylabel={NMR signal (arb. u.)},
		xmin=0, xmax=410,
		minor x tick num=1,
		minor y tick num=1,
		ylabel near ticks, xlabel near ticks,
		]
		\addplot[black, only marks, opacity=0.8, mark= o] table [y expr=\thisrowno{1}/10^5]{./data/exportIrradTime.dat};
	\addplot[red, no markers, line width= 1.0] table [y expr=\thisrowno{1}/10^5]{./data/exportIrradTimeFit.dat};
		\node[red] at (axis cs: 140,0.40) {$T_{\text{pump}}=\SI{102\pm 14}{\second}$};
		\node[black] at (axis cs: 20,2.40) {(a)};
		\end{axis}
		\begin{axis}[xshift=.16\textwidth,yshift=0.90cm,width=0.45\linewidth,font=\scriptsize, ymin=0, ymax=600, axis background/.style={fill=white}, minor x tick num=1,,xlabel={time (\si{\second})},ylabel={FWHM (\si{\hertz})}, ylabel near ticks, xlabel near ticks,height=3.0cm, legend columns=-1]
		\addplot[black, only marks, opacity=0.8, mark= o,restrict x to domain=85:400] table[] {./data/pupmingLinInterpolfwhm.dat};	
		\addlegendentry{raw};
		\addplot[red, only marks, opacity=0.8, mark= o,restrict x to domain=85:400] table[] {./data/pupmingFitfwhm.dat};
		\addlegendentry{fit};
		\end{axis}
		\begin{axis}[
		at={($(plot1.east)+(0.2cm,0)$)},anchor=west,
		height=5.0cm, width=0.3\linewidth,
		xlabel={$B$ (\si{\milli\tesla})},
		ylabel={NMR signal (arb. u.)},
		xmin=49.3, xmax=49.7,
		xtick={49.5},
		ymin=-1.5, ymax=2.25,
		minor y tick num=1, xlabel near ticks,ylabel near ticks, yticklabel pos=right,
		]
		\addplot[black, mark = o, opacity=0.8,] table[y expr=\thisrowno{1}/10^5]{./data/hyperpol_smallScan.dat};
		\node[circle,draw=red, minimum size=10pt, line width=1.0] () at (axis cs: 49.55,1.84) {};
		\node[black] at (axis cs: 49.44,2.0) {(b)};
		\end{axis}
		\end{tikzpicture}
\caption{(a) Integral intensity of the real part of the Fourier transformed NMR signal after different illumination times (\SI{5}{\watt}, \SI{532}{\nano\meter}). The inset shows the corresponding linewidths of the spectra. All values are the average over 3 measurement runs. (b) Section of the magnetic field dependent hyperpolarization pattern. The red circle indicates the magnetic field, where the time dependent measurements were performed.}
\label{pumpingtime}
\end{figure}
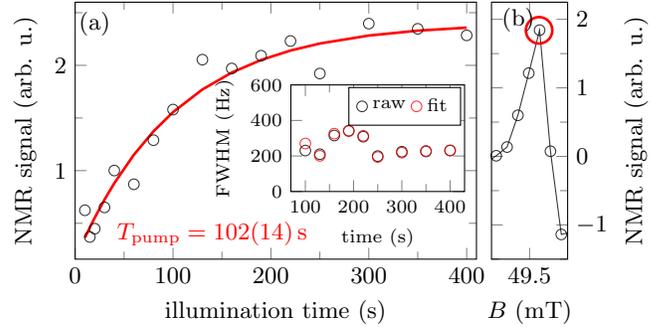
\begin{figure*}[bt!]
\begin{center}
\tikzsetnextfilename{decayDynamics}
\begin{tikzpicture}
\begin{groupplot}[group style={group size=4 by 1},height=4.5cm,width=0.26*\textwidth,minor y tick num=1, ]
\nextgroupplot[xlabel= time (s), ylabel={NMR signal (arb. u.)}, ylabel near ticks, xlabel near ticks, title= below resonances, title style={yshift=-.5em},]
\addplot[black, only marks, opacity=0.8, mark= o] table [y expr=\thisrowno{1}/10^5]{./data/exportLowFieldDecay481G.dat};
\addplot[red, no markers, line width= 1.0] table [y expr=\thisrowno{1}/10^5]{./data/exportLowFieldDecay481GFitnoOffset.dat};
\node at (68pt,68pt) {(a)};
\node at (35pt,65pt) {\SI{48.1}{\milli\tesla}};
\node[red] at (axis cs: 350,0.5) {$T_{1}=\SI{53\pm 5}{\second}$};
\nextgroupplot[xlabel= time (s), xlabel near ticks , title= at resonance, title style={yshift=-.5em}]
\addplot[black,only marks, opacity=0.8 , mark= o] table [y expr=\thisrowno{1}/10^5]{./data/exportLowFieldDecay.dat};
\addplot[red, no markers, line width= 1.0] table [y expr=\thisrowno{1}/10^5]{./data/exportLowFieldDecayFitnoOffset.dat};
\node at (68pt,68pt) {(b)};
\node at (35pt,65pt) {\SI{49.4}{\milli\tesla}};
\node[red] at (axis cs: 135,0.75) {$T_{1}=\SI{70\pm 8}{\second}$};
\nextgroupplot[xlabel= time (s), xlabel near ticks ,title= above resonances, title style={yshift=-.5em}]
\addplot[black, only marks, opacity=0.8, mark= o] table [y expr=\thisrowno{1}/10^5]{./data/exportLowFieldDecay549G.dat};
\addplot[red, no markers, line width= 1.0] table [y expr=\thisrowno{1}/10^5]{./data/exportLowFieldDecay549GFitnoOffset.dat};
\node at (68pt,68pt) {(c)};
\node at (35pt,65pt) {\SI{54.9}{\milli\tesla}};
\node[red] at (axis cs: 350,0.3) {$T_{1}=\SI{55\pm 9}{\second}$};
\nextgroupplot[xlabel= time (min), xlabel near ticks, title= way above resonances,title style={yshift=-.5em}]
\addplot[black, only marks, opacity=0.8, mark= o] table [y expr=\thisrowno{1}/10^5]{./data/exportTimeUltraLong.dat};
\addplot[red, no markers, line width= 1.0] table [y expr=\thisrowno{1}/10^5]{./data/exportTimeUltraLongFit.dat};
\node at (68pt,68pt) {(d)};
\node at (35pt,65pt) {\SI{7.05}{\tesla}};
\node[red] at (axis cs: 270,0.9) {$T_{1}=\SI{139\pm 74}{\minute}$};
\end{groupplot}
\end{tikzpicture}
\caption{Specific decays of the hyperpolarization signal depending on the magnetic field after an illumination time of \SI{200}{\second} (a,c) and \SI{250}{\second} (b,d), respectively (\SI{532}{\nano\meter}, \SI{5}{\watt}). The polarization procedure for each experiment takes place at the indicated field point in Fig.\ref{pumpingtime}\,b, where the pumping time was investigated. The $T_1$ time in the \SI{7.05}{\tesla} field (c) is about \SI{2,5 \pm 1}{\hour} and were measured with the same polarization conditions like in (b).}
\label{timebehavior}
\end{center}
\end{figure*}
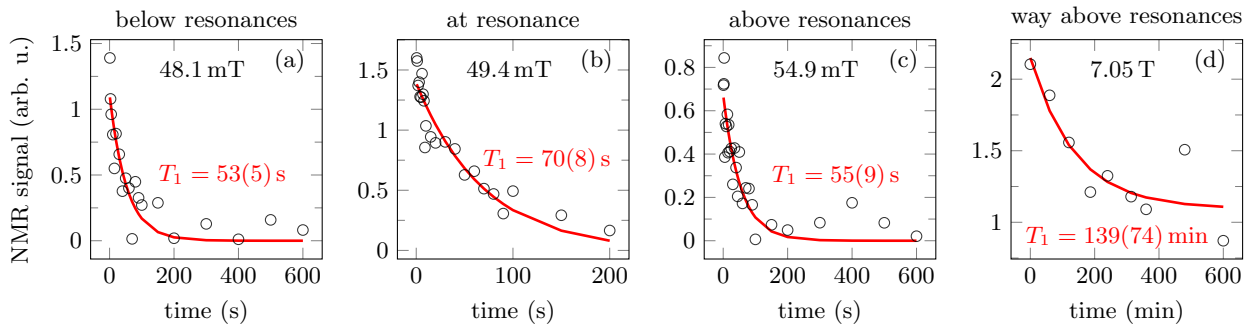
For the build up time measurement, the sample was exposed to a \SI{5}{\watt} laser light at \SI{532}{\nano\meter} for varying illumination times at a magnetic field of about \SI{49.5}{\milli\tesla}. This corresponds to the magnetically first resonant CR between parallel aligned NV centers and P1 centers which are oriented in an angle of \SI{109}{\degree} (Fig.~\ref{pumpingtime}\,b).
After the illumination the sample was transferred into the NMR probe at \SI{7.05}{\tesla} and a $\pi / 2$ pulse was applied immediately. Figure~\ref{pumpingtime}\,a shows the average over three runs of this procedure per illumination time. The characteristic pumping time in this experiment was determined to about $T_{\text{pump}}=\SI{102\pm 14}{\second}$.
In addition, the linewidth (full width at half maximum (FWHM)) of the Fourier transformed NMR signal was analyzed and the values are given in the inset of Fig.~\ref{pumpingtime}\,a and scatter around \SI{200}{\kilo\hertz}. Due to low signal to noise (SNR) ratios for short illumination times, two types of evaluations were performed: first, the FWHM value of the linearly interpolated raw data and second the value extracted from a fit of two Gaussian (for details see Supplemental Material \cite{suppl}).\par
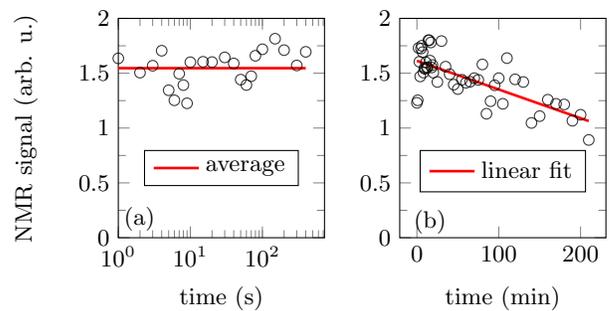
\begin{figure}[bt]
\begin{center}
\tikzsetnextfilename{highFieldDecayShort}
\begin{tikzpicture}
\begin{groupplot}[group style={group size=2 by 1},height=4.5cm,width=0.5\linewidth, ymin=0 ,ymax=2 , minor y tick num=1]
\nextgroupplot[xmin=1, xmode=log,xlabel= time (s), ylabel={NMR signal (arb. u.)},legend style={at={(0.5,0.4)}, anchor=north,legend columns=1},]
\addplot[red, no markers, line width= 1.0] table [y expr=\thisrowno{1}/10^5]{./data/exportHighFieldDecayShortFit.dat};
\addlegendentry{average}
\addplot[black, only marks, opacity=0.8, mark= o] table [y expr=\thisrowno{1}/10^5]{./data/exportHighFieldDecayShort.dat};
\node[black] at (8pt,7pt) {(a)};
\nextgroupplot[xlabel= time (min),legend style={at={(0.5,0.4)}, anchor=north,legend columns=1},]
\addplot[red, no markers, line width= 1.0] table [y expr=\thisrowno{1}/10^5]{./data/exportHighFieldDecayLongFit.dat};
\addlegendentry{linear fit}
\addplot[black, only marks, opacity=0.8, mark= o] table [y expr=\thisrowno{1}/10^5]{./data/exportHighFieldDecayLong.dat};
\node[black] at (5pt,6pt) {(b)};
\end{groupplot}
\end{tikzpicture}
\caption{Time behavior of the hyperpolarization effect at the \SI{7.05}{\tesla} field within the first \SI{400}{\second} (a) and first \SI{200}{\minute} (b) after shuttling. In both cases each data point is averaged over three measurements.}
\label{HighFieldDecay}
\end{center}
\end{figure}

Another important parameter is the typical depletion time of the hyperpolarization signal. For this reason, we measure the decay time at four characteristic magnetic fields, namely slightly below, exactly at, slightly above and far above the resonant spin polarization transfer magnetic field.
The polarization procedure for each experiment takes place at the indicated field point in Fig.\ref{pumpingtime}\,b, where also the pumping time was investigated. The decay of the polarization was measured with a pumping time of \SIrange{200}{250}{\second} which is sufficient to be in the saturated region of the pumping process.
Afterwards, the laser was switched off and the magnetic field was set to the selected value. The NMR measurement takes place after varying duration in this selected magnet field.
The decay was fitted to an single exponential function.
Within the accuracy of the measurement no significant change in the decay times was noticeable.
The values range from \SI{50}{\second} to \SI{80}{\second}, in consideration of the uncertainty of the fits.
This is about \num{2.5} times faster than the characteristic pumping time. A possible reason is discussed below. 
With increasing the magnetic field to \SI{7.05}{\tesla} there is a tremendous increase in the decay time by a factor of \num{200} with a time constant in the range of \SI{2.5}{\hour}.
Due to this long time, no change in the signal can be recognized within in the first \SI{300}{\second} (Fig.~\ref{HighFieldDecay}\,a) and even an observation time over \SI{200}{\minute} (Fig.~\ref{HighFieldDecay}\,b) identifies just a slow decrease of the signal. This decrease appears to be linear due to $\exp{(-t)} \approx 1-t$.
The extreme long $T_1$ time might be a manifestation of the wide off-resonant Zeeman splittings of the two defect systems and may be an advantage for future developments and novel applications.\par
\begin{figure}[bt]
\centering
\begin{minipage}[]{0.60\linewidth}
	\tikzsetnextfilename{SolidEcho}
	\begin{tikzpicture}
	\begin{axis}[
	height=4.5cm, width=1.00\linewidth,
	xlabel={time (\si{\milli\second})},
	ylabel={NMR signal (arb. u.)},
	xmin=0, xmax=5,
	minor x tick num=1,
	ymin=-1.5, ymax=1.25,
	minor y tick num=1,
	xlabel near ticks,
	ylabel near ticks,
	legend style={at={(0.5,0.35)}, anchor=north,legend columns=1},
	]	
	\addplot[gray, opacity = 0.8, line width= 0.5] table {./data/solidecho.dat};
	\addlegendentry{signal};
	\addplot[red, opacity = 1.0, line width= 1.0] table {./data/solidechoFilter.dat};
	\addlegendentry{low pass filtered};
	\end{axis}
	\end{tikzpicture}
\end{minipage}
\begin{minipage}[]{0.38\linewidth}
\centering
\raisebox{\depth}
{\begin{tabular}{ccc}
$\pi/2$ & $\pi/2_{90\si{\degree}}$ & receiver\\\hline
0	&	1	&	0\\
0	&	3	&	0\\
1	&	0	&	1\\
1	&	2	&	1\\
2	&	1	&	2\\
2	&	3	&	2\\
3	&	0	&	3\\
3	&	2	&	3\\\\\\
\end{tabular}}
\end{minipage}
\put(-200,46){(a)}
\put(-102,46){(b)}
\caption{(a) The signal of a hyperpolarized solid echo experiment with $\tau = \SI{500}{\micro\second}$ in the time domain. (b) The table shows the phases of one phase cycle for the first and second $\pi/2$ pulse as well as the receiver phase. Here $0$, $1$, $2$, $3$ correspond to a phase of $0$, $\pi/2$, $\pi$ and $3\pi/2$. The signal in (a) is accumulated over two phase cycles.}
\label{solidecho}
\end{figure}
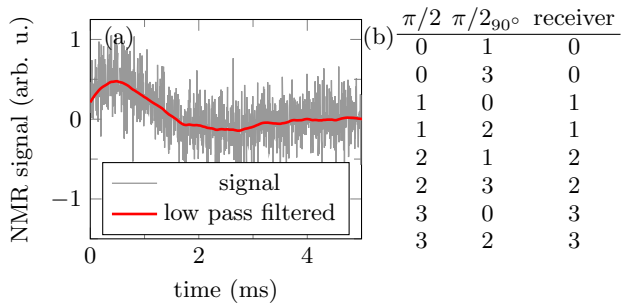

For a prove of concept and as reliability check a series of solid echo measurements was conducted under hyperpolarization. This kind of pulse sequence is used to verify dipolar coupling of magnetically equivalent  spin-1/2 pairs among themselves.
After a specific delay time (in this case \SI{0.5}{\milli\second}) the first $\pi/2$ pulsed is followed by a second one with a relative phase of \SI{\pm 90}{\degree}. The receiver phase is equal to that of the first pulse.
The accumulated signal in the time domain over two full phase cycles (8 measurements per cycle) is shown in Fig.~\ref{solidecho}\,a and the corresponding pulse sequence is given in the table (Fig.~\ref{solidecho}\,b). The clear increase after \SI{0.5}{\milli\second} indicates a dipolar coupling of the hyperpolarized $^{13}$C spin.
\FloatBarrier
\subsection{Thermal equilibrium NMR}
In addition to the hyperpolarization measurements, conventional measurements in quasi thermal (QT) equilibrium were conducted in the very same setup. This means, the diamond was attached to the transfer shuttle but stays in the NMR probe in the center of the 7.05-\si{\tesla} magnet.
The absence of hyperpolarization and the long lattice relaxation time requires a large number of accumulations and causes a long measurement times.
Standard FID measurements with different delay times between every sequence were performed. The linewidths of the NMR spectra are calculated as described above and shown in Fig.~\ref{300er} (see Supplemental Material for raw data). 
With increasing delay time, the linewidth decreases and reaches a value around \SI{300}{\hertz} for a delay time of \SI{4}{\hour}.\par
\begin{figure}[bt]
\centering
\tikzsetnextfilename{Thermal300er}
\begin{tikzpicture}
\begin{axis}[xmode=linear, xlabel={time (\si{\minute})}, ylabel={FWHM (\si{\hertz})},legend style={at={(0.8,0.9)}, anchor=north,legend columns=1},  height=4.0cm,width=1.0\linewidth,xlabel near ticks,  ylabel near ticks,]
\addplot[black, opacity=0.8, mark= o] table[x expr=\thisrowno{0}*1] {./data/300erLinInterpolfwhm.dat};	
\addlegendentry{raw};
\addplot[red, opacity=0.8, mark= o] table[x expr=\thisrowno{0}*1] {./data/300erFitfwhm.dat};
\addlegendentry{fit};
\end{axis}
\end{tikzpicture}
\caption{Extracted linewidths from non hyperpolarized measurements with different repetition times (\SI{11}{\minute}, \SI{30}{\minute}, \SI{1}{\hour}, \SI{2}{\hour}, \SI{4}{\hour}) corresponding to an accumulation of \num{267}, \num{87}, \num{38}, \num{36} and \num{73} single measurements. Here, the repetition time is defined as the time between two measurement sequences, consisting of a $\pi/2$ pulse and a subsequent acquisition of the FID.}
\label{300er}
\end{figure}
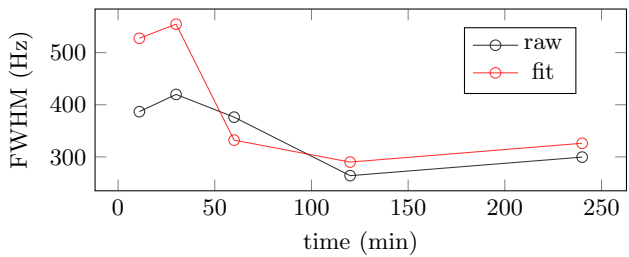
To improve the SNR and to investigate the nuclear spin system in more detail, we performed QT measurement in a 11.74-\si{\tesla} magnet (\SI{500}{\mega\hertz} proton Larmor frequency) in combination with a Bruker Avance III HD spectrometer. The diamond sample was placed in a HF coil with a quality factor of $Q\approx 190$ and was oriented with its crystallographic $[111]$ direction parallel to the applied magnetic field. A carrier frequency of \SI{125,758189}{\mega\hertz} was used and the length for a $\pi/2$ pulse was determined to be \SI{5,5}{\micro\second}.
The measurements were conducted as ``saturation recovery'' with four $\pi/2$ pulses at the beginning of each measurement sequence.
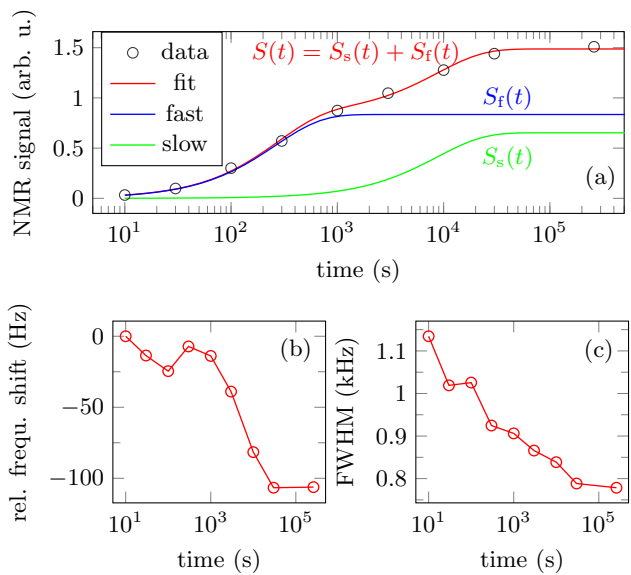
\begin{figure}
\begin{center}
\tikzsetnextfilename{Thermal500erA}
\begin{tikzpicture}
\begin{axis}[xmode=log, xlabel= time (\si{\second}),ylabel={NMR signal (arb. u.)},legend style={at={(0.13,1.00)}, anchor=north,legend columns=1}, xmin=5, xmax=5e5, height=4.0cm,width=1.0\linewidth, xlabel near ticks,  ylabel near ticks,]
\addplot[black, only marks, opacity=0.8, mark= o] table[y expr=\thisrowno{1}/10^5]{./data/export500erT1timeData.dat};
\addlegendentry{data};
\addplot[red, no markers, line width= 0.5] table[y expr=\thisrowno{1}/10^5]{./data/export500erT1timeFit.dat};
\addlegendentry{fit};
\addplot[blue, no markers, line width= 0.5] table[y expr=\thisrowno{1}/10^5]{./data/export500erT1timeFast.dat};
\addlegendentry{fast};
\addplot[green, no markers, line width= 0.5] table[y expr=\thisrowno{1}/10^5]{./data/export500erT1timeSlow.dat};
\addlegendentry{slow};
\node[red] at (axis cs: 1500,1.45) {$S(t) = S_{\text{s}}(t) + S_{\text{f}}(t)$};
\node[green] at (axis cs: 40000,0.4) {$S_{\text{s}}(t)$};
\node[blue] at (axis cs: 40000,1.0) {$S_{\text{f}}(t)$};
\node at (axis cs: 3e5,0.2) {(a)};
\end{axis}
\end{tikzpicture}
\tikzsetnextfilename{Thermal500erBC}
\begin{tikzpicture}
\begin{groupplot}[group style={group size=2 by 1},height=4.0cm,width=0.51\linewidth , minor y tick num=1, xmin=5, xmax=5e5]
\nextgroupplot[,xmode=log, ylabel= rel. frequ. shift (\si{\hertz}),xlabel= time (\si{\second}),xlabel near ticks, ylabel near ticks, ]
\addplot[red, mark=o, line width= 0.5] table {./data/export500erFrequShift.dat};
\node at (axis cs: 100000,-10) {(b)};
\nextgroupplot[xshift=0.2cm,xmode=log,ylabel= FWHM (\si{\kilo\hertz}), xlabel= time (\si{\second}),xlabel near ticks,  ylabel near ticks,]
\addplot[red, mark=o, line width= 0.5] table {./data/export500erFWHM.dat};
\node at (axis cs: 100000,1.1) {(c)};
\end{groupplot}
\end{tikzpicture}
\caption{(a) Integrated NMR signal for different times after saturation at \SI{11.74}{\tesla}. The fitting parameters are: $T^{\text{f}}_1=\SI{263\pm24}{\second}$, $T^{\text{s}}_1=\SI{8700\pm1300}{\second}$ with the ratio $A_{\text{s}} / A_{\text{f}} \approx \num{0.8}$. A shift in the center frequency (b) as well as a narrowing of the NMR peaks (c) with increasing time is clearly visible.}
\label{500er}
\end{center}
\end{figure}
Figure~\ref{500er}\,a shows the increasing integral NMR signal for different delay times after die saturation pulses.
For the first two data points (\SI{10}{\second}, \SI{30}{\second}) \num{512} scans were accumulated, four scans for the data points from \SIrange{e2}{3e4}{\second} and a single measurement for \SI{26e4}{\second} ($\approx \SI{3}{\day}$).
The data is fitted to a double exponential function with a slow $S_{\text{s}}(t)$ and a fast component $S_{\text{f}}(t)$:
\begin{align*}
S(t) &= S_{\text{s}}(t) + S_{\text{f}}(t) \\
&= A_{\text{s}} \cdot \left( 1-\exp\left(-\frac{t}{T^{\text{s}}_1}\right) \right) + A_{\text{f}} \cdot \left( 1-\exp\left(-\frac{t}{T^{\text{f}}_1}\right) \right) .
\label{T1thermalFit}
\end{align*}
The fit yields $T^{\text{s}}_1=\SI{2.0\pm 0.4}{\hour}$ and $T^{\text{f}}_1=\SI{4.4\pm 0.4}{\minute}$, respectively.\par
A closer look at the peaks reveals a shifting of the center frequency as well as a narrowing of the peak width (Fig.~\ref{500er}\,b,c) with increasing delay time. Although the effect of shifting is rather small the line width (FWHM) decreases by \SI{30}{\percent} from \SI{1.1}{\kilo\hertz} to \SI{0.8}{\kilo\hertz}.
This corroborates the trend of a decreasing linewidth with increasing repetition time of the QT measurement at \SI{7.05}{\tesla}.
\section{Discussion}
The results and in particular the comparison of hyperpolarized and QT measurements give evidence that the described method of CR is based on weakly coupled nuclear spins and induces spin diffusion. This is discussed further in the following section.\par
\subsection{Magnetic field dependence}
First of all, the distance of \SI{0.114}{\milli\tesla} of each peak pair of the magnetic field sweep corresponds to a coupling of the $^{13}$C spins in the range of \SI{2}{\mega\hertz} \cite{Wunderlich2017,Pagliero2018}.
At this point, it is unclear if this is intrinsically caused by the hyperpolarization process or can be explained by statistical arguments. Since the strongly coupled $^{13}C$ spins in the first shells around the paramagnetic center have a lower probability of occurrence. The first shell hyperfine (hf) coupling parameters for a NV center in the principal axis system are $A_{xx}=\SI{30}{\mega\hertz}$, $A_{yy}=\SI{123}{\mega\hertz}$ and $A_{zz}=\SI{227}{\mega\hertz}$ \cite{Shim2013}. Therefore, this coupling can not explain the experimental data.
In reference~\citenum{Smeltzer2011} it was found experimentally as well as verified by theoretical ab into calculations, that a NV-$^{13}$C hf coupling around \SI{2.5}{\mega\hertz} can be associated with 9 possibles sites in a distance of \SI{5}{\angstrom}. In Ref.~\citenum{Dreau2012} couplings even below \SI{1}{\mega\hertz} are reported.
For P1 centers hf couplings with $^{13}$C spins bewtween \SI{341}{\mega\hertz} and \SI{1}{\mega\hertz} depending on the lattice site are reported \cite{Barklie1981,Cox1994,Peaker2016}.
Hence, the assumed hf coupling of $\sim \SI{2}{\mega\hertz}$ between the regarded paramagnetic centers and a $^{13}$C spin is in accordance with the current literature.\par
\subsection{Thermal measurements}
Taking into account the double exponential decay as well as the narrowing of the line width in the QT measurements at \SI{11.74}{\tesla}, the $^{13}$C spins can be separated at least in two groups. The first one is located in the neighborhood of paramagnetic defects like NV or P1 centers (fast decay with $T^{\text{f}}_1=\SI{4.4\pm 0.4}{\minute}$, broad line width) and the second group far away from any paramagnetic impurities (slow decay with $T^{\text{s}}_1=\SI{2.0\pm 0.4}{\hour}$, narrow line width).
Regarding the hyperpolarization decay at \SI{7.05}{\tesla} for short times, like shown in Fig.\ref{HighFieldDecay}\,a, no change in the NMR intensity in the range of $T^{\text{f}}_1= \sim \SI{4}{\minute}$ is noticeable. Obviously, the main part of the hyperpolarized signal is contributed by $^{13}$C  beyond a minimal distance to the paramagnetic defects with long $T_1$ times. This is emphasized by the fact, that the line width of the hyperpolarized data is comparable with the QT measurements at \SI{7.05}{\tesla} only for long delay times in the latter (Fig.~\ref{300er}).\par
\begin{figure}[bt]
\centering
\tikzsetnextfilename{noisedensity}
\begin{tikzpicture}
\begin{axis}[xmode=log, xmin=0.01, xmax=300, ymin=-0.1, ymax= 1.1, samples=500, xlabel=frequency (\si{\mega\hertz}), ylabel= norm. noise, width=\linewidth, height=5cm, xlabel near ticks,  ylabel near ticks, legend style={at={(1.00,1.2)}},legend columns=2, ]
  \addplot[name path=plot11, red,forget plot][domain=0.01:1] (x,{1.0*(1/(1+(1*x)^2))});
  \addplot[name path=plot21, red,forget plot][domain=0.01:1]  (x,{0.04*(25/(1+(25*x)^2))});
  \addplot[name path=plot12, red,forget plot][domain=1:300] (x,{1.0*(1/(1+(1*x)^2))});
  \addplot[name path=plot22, red,forget plot][domain=1:300]  (x,{0.04*(25/(1+(25*x)^2))});
  \addplot[thick, color=red, fill=red, fill opacity=0.5] fill between[of = plot11 and plot21,soft clip={domain=0.01:1}];
  \addplot[thick, color=red, fill=red, fill opacity=0.5, forget plot] fill between[of = plot12 and plot22,soft clip={domain=1:300}];    
  \addlegendentry{$\tau_{\text{c}}=$ \SIrange{1}{25}{\micro\second}};
  \addplot[blue][domain=0.01:10]  (x,{10*(0.1/(1+(0.1*x)^2))});
  \addlegendentry{$\tau_{\text{c}}=\SI{0.1}{\micro\second}$};
  \addplot[blue][domain=10:300]  (x,{10*(0.1/(1+(0.1*x)^2))});
  \addplot[gray, line width = 5,opacity=0.5] (0.535,x);
  \addplot[gray, line width = 5,opacity=0.5] (75.4689,x);
  \addplot[gray, line width = 5,opacity=0.5] (125.7582,x);
  \node[rotate=90] at (axis cs: 0.535,0.9) {\SI{50}{\milli\tesla}};
  \node[rotate=90] at (axis cs: 75.4689,0.9) {\SI{7.05}{\tesla}};
  \node[rotate=90] at (axis cs: 125.7582,0.9) {\SI{11.74}{\tesla}};
\end{axis}
\end{tikzpicture}
\caption{Normalized spectral noise density assuming a exponentially decaying auto correlation function for different correlation times $\tau_{\text{c}}$. The gray regions indicate the $^{13}$C Larmor frequencies for the three used magnetic fields of the presented measurements.
}
\label{noise}
\end{figure}
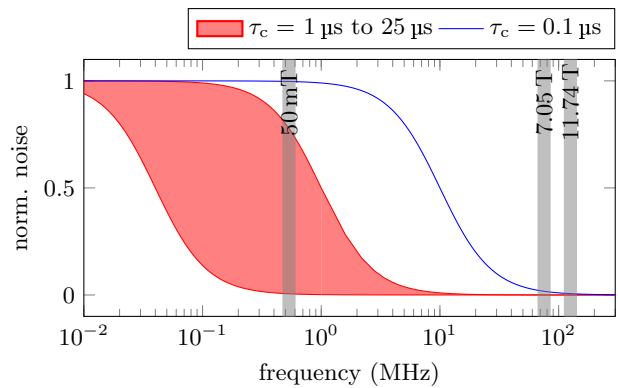
\subsection{Spin diffusion}
The dipolar echo experiment indicates a dipolar coupling among the hyperpolarized $^{13}$C spins, which is a requirement for spin diffusion.
But spin diffusion can take place only if the Larmor frequency of neighboring spins are in resonance. In the vicinity of a paramagnetic center a strongly magnetic field gradient is produced, inducing a shift of the nuclear Larmor frequency of the surrounding spins depending on their distance to the defect. This creates a diffusion barrier in the distance $b$ around the defect. Within a radius $r<b$ the diffusion is supressed (diffusion constant $D=0$). The increase of $D$ to its unperturbed value can be described by $exp[-(b/r)^8]$ \cite{Khutsishvili1966}. Assuming a hf coupling of $\sim \SI{2}{\mega\hertz}$ for NV-$^{13}$C and extracting the NV-$^{13}$C distance from Ref.~\citenum{Smeltzer2011} results in $b=\SI{5}{\angstrom}$. The nuclear hf coupling with P1 centers is highly anisotropic but for a distance of \SI{2.6}{\angstrom} values from \SIrange{1}{3}{\mega\hertz} are found \cite{Peaker2016}.
Using the equation from Ref.~\citenum{Terblanche2001} the diffusion barrier radius can be estimated with
\begin{equation*}
b=\left( \frac{\hbar \gamma_e^2 B}{2 k_{\text{B}} T \gamma_{\text{13C}}}\right)^{1/4} \cdot a
\end{equation*}
and gives $b=\SI{3.2}{\angstrom}$ for the P1 center. This corresponds roughly to an exclusion of $^{13}$C~spins in the range of one lattice constant.
Here, $\gamma_e$ denotes the electronic gyromagnetic ratio, $\gamma_{\text{13C}}$ the $^{13}$C gyromagnetic ratio, $k_{\text{B}}$ the Boltzmann constant, $T$ the temperature, $B$ the external magnetic field and $a$ the average nearest neighbor $^{13}$C distance, with  $T=\SI{300}{\kelvin}$ and $a=\SI{4.4}{\angstrom}$. The parameter $a$ is calculated via $1/2=\exp{\lbrace-(4\pi N_{\text{13C}} a^3 /3)\rbrace}$ to $a=0.55N_{\text{13C}}^{-1/3}$ \cite{Reynhardt2001}. This is depicted in the sketch of Fig.~\ref{T1noise}. For the NV center an additional region around the defect exists, where the strongly coupled $^{13}$C spins get hyperpolarized directly via the excited state level anticrossing (ESLAC) in the investigated field region \cite{Smeltzer2009}. On the one hand this process seems to be less efficient for the bulk hyperpolarization and on the other hand this ESLAC polarized region overlaps at least partly with the diffusion barrier region.\par
The nuclear spin diffusion itself, shows a low diffusion constant  $D=\SI{67}{\square\angstrom\per\second}$ leading to a slow propagation and therefore a short spatial range (see Supplemental Material) \cite{Terblanche2001}. However, the polarization in the order of several percent indicates a spread of the polarization over a wide region in the sample. This can not be explained by classical $^{13}$C spin diffusion solely. An unknown effect seems to increase the diffusion range. We speculate that the diffusion enhancement is driven by the dipolar coupled P1 network.
\subsection{Time dependence}
The characteristic pumping time is at least roughly twice the decay time in the same field region.
On the one hand, an unstable laser output in the first seconds can not be excluded, causing the longer pumping time. On the other hand, a reason for this could be the spread of the hyperpolarization over the bulk via diffusion and the fact, that the NV system is frequently in the excited state during the polarization. But the upper-state electronic configuration is different from that in the ground state, where the NV center is in resonance with the P1 defects.\par
\begin{figure}[bt]
\centering
\tikzsetnextfilename{noiseT1}
\begin{tikzpicture}
\begin{axis}[xmode=linear, xmin=0.5, xmax=3.5, ymin=40, ymax=100, samples=100, xlabel=correlation time $\tau_{\text{c}} (\si{\micro\second})$ , ylabel= $T_1$ (\si{\second}), width=1.0\linewidth, height=5cm, xlabel near ticks,  ylabel near ticks, legend style={at={(1.00,1.2)}},legend columns=1, ]
  \addplot[red][domain=0.5:3.5] (x,{1.0/((2.0e-2)*(x/(1+(0.532*x)^2)))});
  \addlegendentry{$f_{\text{L}} (\SI{50}{\milli\tesla})=\SI{0,532}{\mega\hertz}$};
  \addplot +[gray, line width = 5,opacity=0.5, mark=none] coordinates {(1.88,40)(1.88,100)};
  \node[rotate=90] at (axis cs: {1/0.532},90) {$1/f_{\text{L}}$};
\end{axis}
\fill[red] (1.5+3.5,1.5+0.25) circle(0.1cm);
\draw[black, dashed] (1.5+3.5,1.5+0.25) circle(0.5cm);
\draw[black, dashed] (1.5+3.5,1.5+0.25) circle(1cm);
\draw[black, dashed] (1.5+3.5,1.5+0.25) circle(1.4cm);
\draw[->] (1.5+3.5,1.5+0.25) -- (2.0+3.5,1.5+0.25);
\draw[->] (1.5+3.5,1.5+0.25) -- (2.207+3.5,2.207+0.25);
\draw[->] (1.5+3.5,1.5+0.25) -- (1.5+3.5,2.9+0.25);
\node at (1.75+3.5,1.35+0.25) {$b$};
\node at (2.2+3.5,1.95+0.25) {$p$};
\node at (1.6+3.5,2.7+0.25) {$d$};
\end{tikzpicture}
\caption{Dependence of the $^{13}$C relaxation time $T_1$ from the correlation time $\tau_{\text{c}}$. The vertical gray line indicates the Larmor period of $^{13}$C spins in a magnetic field of around \SI{50}{\milli\tesla}. The red line corresponds to the equation \ref{T1correlationtimes} using a $\langle B_x^2
\rangle = \SI{2e-4}{\square\tesla}$. The inset depicts the different spheres around a paramagnetic defect (red). The parameter $b$ denotes the characteristic radius of the diffusion barrier, $p$ the radius for nuclear polarization via CR and $d$ the diffusion radius.}
\label{T1noise}
\end{figure}
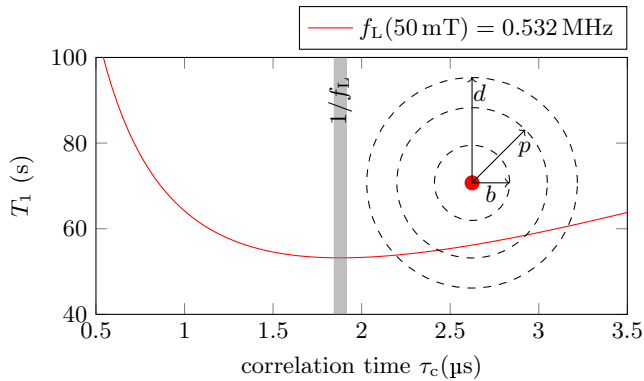
Besides the $^{13}$C spins, the NV centers as well as the coupled P1 centers are polarised during the laser illumination \cite{Loretz2017}. After switching off the laser light, the electronic spins of the paramagnetic centers decay quickly back into thermal equilibrium and induce magnetic noise which can influence the T1 time of the nuclear $^{13}$C spins.
In the literature, the typical correlation times $\tau_{\text{c}}$ for magnetic noise around NV centers ranges from \SIrange{1}{25}{\micro\second} \cite{Bar-Gill2012,Wang2012,Lange2012}. For example, in Ref.~\citenum{Bar-Gill2012} a  $\tau_{\text{c}}=\SI{3\pm 2}{\micro\second}$ for a nitrogen concentration of \SI{100}{ppm} and a NV density of \SI{e16}{\per\square\centi\meter} were found.
Assuming a Lorentzian spectral density for this noise, this can effect the relaxation times for the $^{13}$C spins around \SI{50}{\milli\tesla} but not in the high field regime at \SI{7.05}{\tesla} and \SI{11.74}{\tesla}. Even a correlation time of $\tau_{\text{c}}=\SI{0.1}{\micro\second}$ would have a low spectral density in the high field region (Fig.~\ref{noise}). Lower values of $\tau_{\text{c}}$ are reported for very impure systems like surface near NV centers in nanodiamonds \cite{Song2014}.
A correlation time for the magnetic noise of $\tau_{\text{c}} > \SI{0.1}{\micro\second}$  would explain the unchanged decay time of the hyperpolarization at the three measured low magnetic fields and the long relaxation time for high magnetic fields. For a random fluctuating magnetic field $B_x$ this can be modeled by the formula
\begin{equation}
\frac{1}{T_1} = \gamma_{\text{13C}} \langle B_x^2 \rangle \frac{\tau_c}{1+(\omega_L\tau_c)^2},
\label{T1correlationtimes}
\end{equation}
following from the theory of random fluctuating magnetic fields \cite{LevittBook}. According to this equation, the previous reported correlation times around \SI{3}{\micro\second} give reasons for the observed $T_1$ times in the low magnetic field region (Fig.~\ref{T1noise}).
\begin{figure}[bt]
\centering
\tikzsetnextfilename{PolarisSpheres}
\begin{tikzpicture}[background rectangle/.style={fill=white}, show background rectangle]
\begin{sphere}[sphere scale=4, sphere color=cyan]
    \begin{sphere}[sphere scale=2, sphere color=green]
        \begin{sphere}[sphere scale=1, sphere color=red]
 			\draw [thick,->, color=green, opacity = 1] (0-0.2,0-0.4) -- (0+0.2,0+0.4);
 			\shade [ball color = green, opacity = 1] (0,0) circle (0.25);
        \end{sphere}
     \draw[green, decoration = {zigzag,segment length = 0.5mm, amplitude = 0.5mm},decorate] (0,0)--(1,0.75);
     \draw[blue, decoration = {zigzag,segment length = 1mm, amplitude = 0.5mm},decorate] (1,0.75)--(2,1.25);
     \draw[blue, decoration = {zigzag,segment length = 1mm, amplitude = 0.5mm},decorate] (2,1.25)--(3,2.0);
	 	\draw [thick,->, color=blue, opacity = 1] (3,2-0.2) -- (3,2+0.2);
     \shade [ball color = blue, opacity = 1] (3,2) circle (0.1);
     	 \draw [thick,->, color=blue, opacity = 1] (2,1.25-0.2) -- (2,1.25+0.2);
     \shade [ball color = blue, opacity = 1] (2,1.25) circle (0.1);
     	 \draw [thick,->, color=blue, opacity = 1] (1,0.75-0.2) -- (1,0.75+0.2);
     \shade [ball color = blue, opacity = 1] (1,0.75) circle (0.1);
     	 \draw [thick,->, color=blue, opacity = 1] (-1.5,-2-0.2) -- (-1.5,-2+0.2);
     \shade [ball color = blue, opacity = 1] (-1.5,-2) circle (0.1);
     	 \draw [thick,->, color=blue, opacity = 1] (-1,-1-0.2) -- (-1,-1+0.2);
     \shade [ball color = blue, opacity = 1] (-1,-1) circle (0.1);
     	 \draw [thick,->, color=blue, opacity = 1] (2,-2-0.2) -- (2,-2+0.2);
     \shade [ball color = blue, opacity = 1] (2,-2) circle (0.1);
     	 \draw [thick,->, color=blue, opacity = 1] (1.75,-1.5-0.2) -- (1.75,-1.5+0.2);
     \shade [ball color = blue, opacity = 1] (1.75,-1.5) circle (0.1);
     	 \draw [thick,->, color=blue, opacity = 1] (0.5,1-0.2) -- (0.5,1+0.2);
     \shade [ball color = blue, opacity = 1] (0.5,1) circle (0.1);
     	 \draw [thick,->, color=blue, opacity = 1] (-1,0.75-0.2) -- (-1,0.75+0.2);
     \shade [ball color = blue, opacity = 1] (-1,0.75) circle (0.1);
     	 \draw [thick,->, color=blue, opacity = 1] (-2,0-0.2) -- (-2,0+0.2);
     \shade [ball color = blue, opacity = 1] (-2,0) circle (0.1);
     	 \draw [thick,->, color=blue, opacity = 1] (-3,1-0.2) -- (-3,1+0.2);
     \shade [ball color = blue, opacity = 1] (-3,1) circle (0.1);
     	 \draw [thick,->, color=blue, opacity = 1] (0,-2-0.2) -- (0,-2+0.2);
     \shade [ball color = blue, opacity = 1] (0,-2) circle (0.1);
     	 \draw [thick,->, color=blue, opacity = 1] (-2,2.5-0.2) -- (-2,2.5+0.2);
     \shade [ball color = blue, opacity = 1] (-2,2.5) circle (0.1);
     	 \draw [thick,->, color=blue, opacity = 1] (0.5,-3-0.2) -- (0.5,-3+0.2);
     \shade [ball color = blue, opacity = 1] (0.5,-3) circle (0.1);
     	 \draw [thick,->, color=blue, opacity = 1] (0.75,3.5-0.2) -- (0.75,3.5+0.2);
     \shade [ball color = blue, opacity = 1] (0.75,3.5) circle (0.1);
 	\draw [thick,->, color=yellow, opacity = 1] (-0.9+0.2,-2.8-0.4) -- (-0.9-0.2,-2.8+0.4);
 	\shade [ball color = yellow, opacity = 1] (-0.9,-2.8) circle (0.25);
 	\draw[white, decoration = {zigzag,segment length = 0.5mm, amplitude = 0.5mm},decorate] (0,0)--(-0.9,-2.8);
    \end{sphere}
\end{sphere}
\node[red, fill=lightgray, opacity = 0.7] at (-1,0.2) {barrier};
\node[green, fill=lightgray, opacity = 0.7] at (-2,1.5) {cross relaxation};
        text shadow={[align=center] at (-1,0.2) {barrier}}]
                                      at (-1,0.2) {barrier};
        text shadow={[align=center] at (-2,1.5) {cross relaxation}}]
                                      at (-2,1.5) {cross relaxation};
     \node[cyan,fill=lightgray, opacity = 0.7,
        text shadow={[align=center] at (-3.5,2.75) {spin diffusion}}]
                                      at (-3.5,2.75) {spin diffusion};
\end{tikzpicture}
\caption{Sketch of the hyperpolarization mechanism in the picture of CR between a P1 and NV center (yellow and green spin). The defect is surrounded by a diffusion barrier (red) and a spherical shell of direct coupled $^{13}$C~spins (green). The electronic spin of one defect center is coupled to a nearby $^{13}$C nuclear spin (blue spins), which leads to a polarization transfer to the latter. This polarization can diffuse in the bulk via the dipolar $^{13}$C network. The range of spin diffusion is indicated by the outer sphere.}
\label{spheres}
\end{figure}
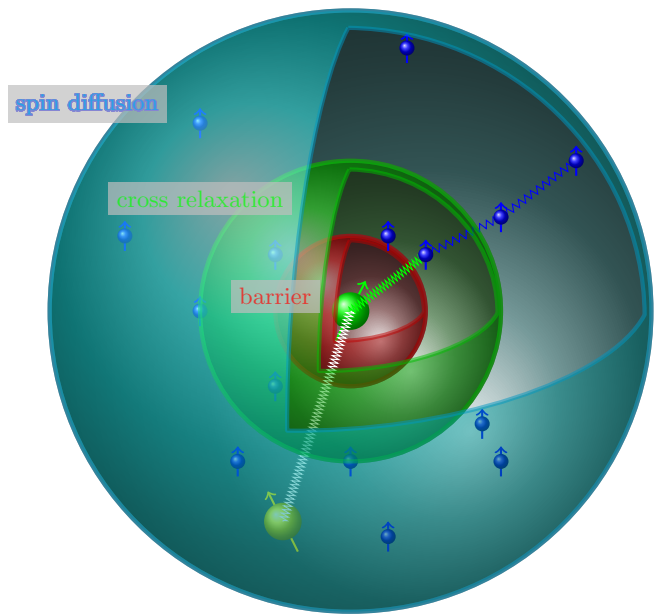
\section{Conclusion and Outlook}
Hyperpolarization measurements were compared with conventional measurements in QT equilibrium.
The linewidth as well as the lattice relaxation times in both cases indicate a weak coupling to the directly hyperpolarized $^{13}$C spins. This assertion is supported by the magnetic field dependent hyperpolarization pattern.
The radius of direct hyperpolarization via CR and the diffusion barrier was estimated using current literature.
A sketch of this model is depicted in Fig.~\ref{spheres}. The coupling between a NV and a P1 center leads via CR to a polarization transfer to a nearby $^{13}$C~spin in a minimal distance beyond the diffusion barrier. This polarization can be passed to the bulk $^{13}$C~spin via the dipolar network.\par
Further studies should also take into account an increasing range of effective spin diffusion with the aid of the network of coupled P1 center. This could enhance the effective diffusion distance due to the stronger electronic dipolar coupling by a factor of $\gamma_{\text{P1}}/\gamma_{\text{13C}} \approx \num{2600}$.\par
Similar investigations of samples with varying defect concentrations could deliver valuable information of this hyperpolarization mechanism. This can open a way towards tailoring desired enhancement factors and relaxation times by defining the average defect distances.
%
\begin{acknowledgments}
This work was supported by the VolkswagenStiftung. We thank Dr. W. Knolle (Leibniz Institute of Surface Engineering (IOM), Leipzig, Germany) for helpful discussions and valuable assistance during the high energy electron irradiation.
Furthermore, we acknowledge support by P. R\"acke in preparing the manuscript.
\end{acknowledgments}
%
\bibliography{hyperpolTimeDyn}
\end{document}


\widetext
\begin{center}
\textbf{\large Supplemental Materials: Time Dynamics of Optically Induced Cross Relaxation via Nitrogen Related Defects for Bulk Diamond \textsuperscript{13}C Hyperpolarization}
\end{center}
\setcounter{equation}{0}
\setcounter{figure}{0}
\setcounter{table}{0}
\setcounter{page}{1}
\makeatletter
\renewcommand{\theequation}{S\arabic{equation}}
\renewcommand{\thefigure}{S\arabic{figure}}
\renewcommand{\bibnumfmt}[1]{[SR#1]}
\renewcommand{\citenumfont}[1]{SR#1}
%
\FloatBarrier
\section{Hyperpolarization NMR signals and linewidth}
\begin{figure}[h]
\centering
\includegraphics[width=0.49\linewidth]{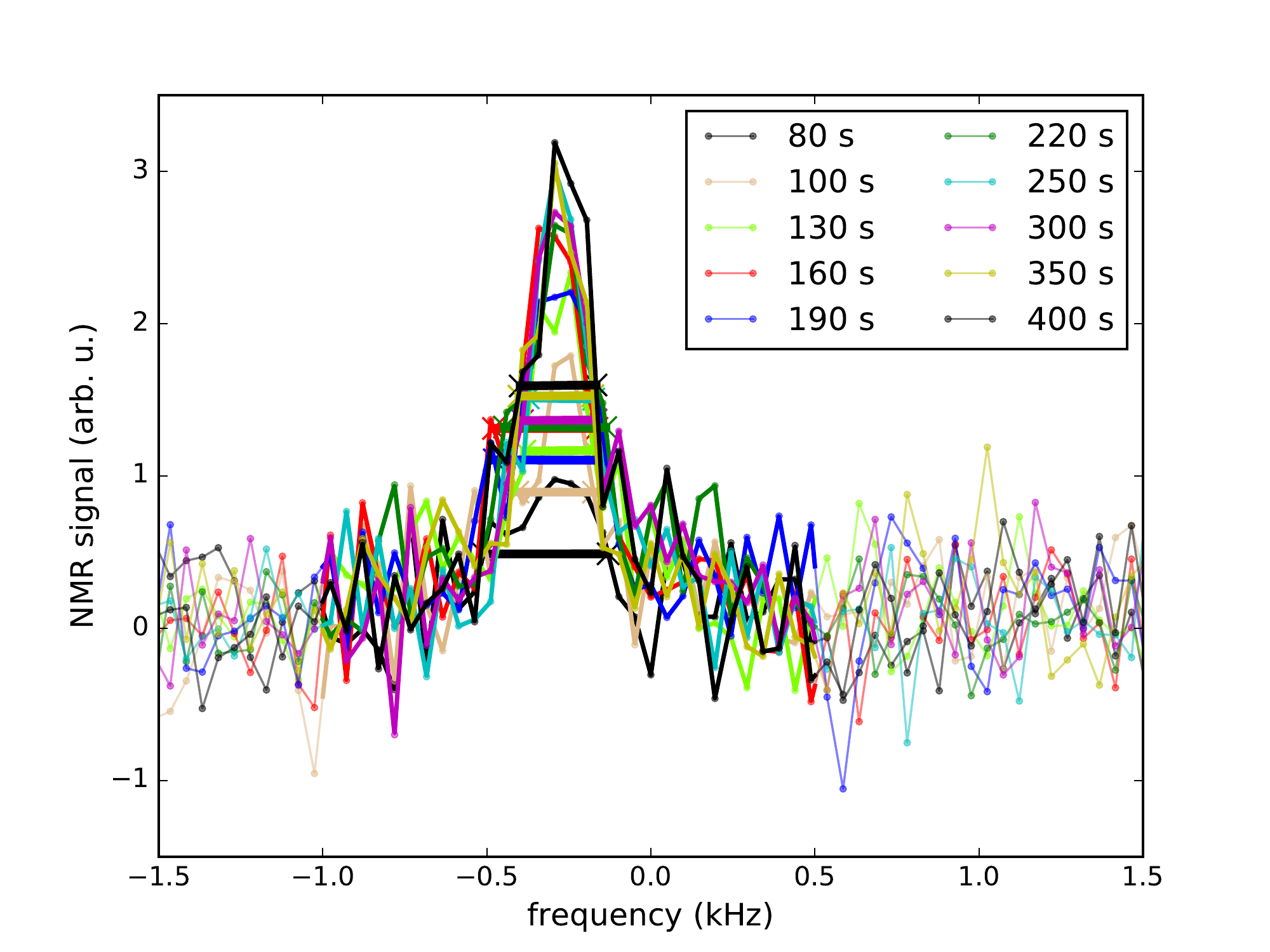}
\includegraphics[width=0.49\linewidth]{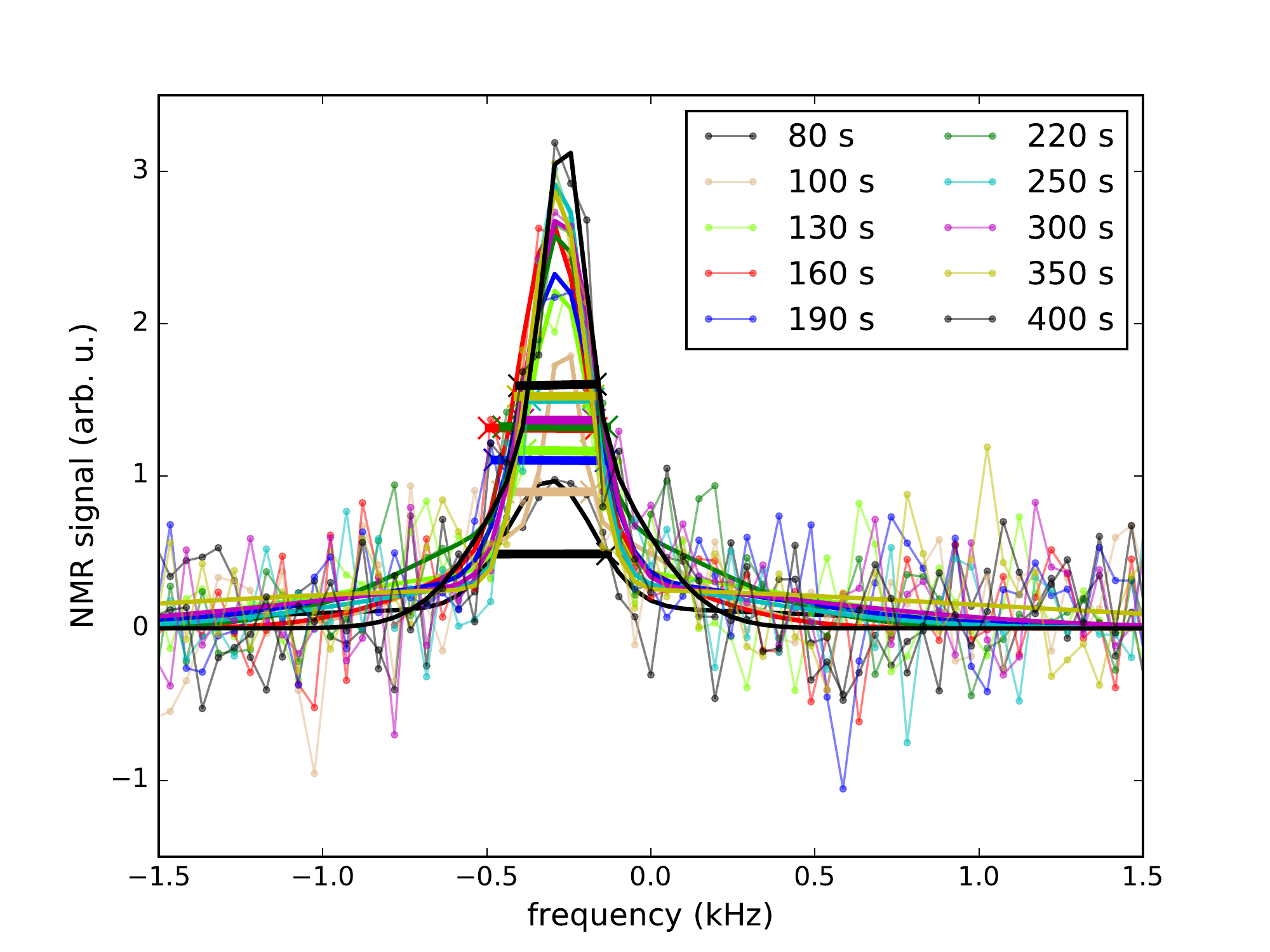}
\put(-465,155){(a)}
\put(-210,155){(b)}
\caption{Fourier transformed NMR signal (real part) after different illumination times. The spectra are normalized on the number of accumulated measurements. Extraction of the linewidth (FWHM) using a linear interpolation (a) as well as a double Gaussian (b).}
\label{pumpingtimeSpectraFWHM}
\end{figure}

Due to noisy NMR spectra two types of linewidth analyses were used. The first one uses a linear interpolation of the raw date (Fig.~\ref{pumpingtimeSpectraFWHM}\,a) and the second one a fitting procedure (Fig.~\ref{pumpingtimeSpectraFWHM}\,b).
The fit function is
\begin{equation}
S(f)=\sum_{i=1}^2 A_i \frac{1}{w_i\sqrt{\pi}} \exp{\left[-\frac{1}{2}\left(\frac{f-f_0}{w_i}\right)^2\right]} + S_0,
\end{equation}
with $A_i$, $f_0$, $w_i$ and $S_0$ as free parameters. The value of the line width is calculated numerically.
\FloatBarrier
\section{Quasi thermal NMR signals and linewidth}
\begin{figure}[h]
\centering
\includegraphics[width=0.49\linewidth]{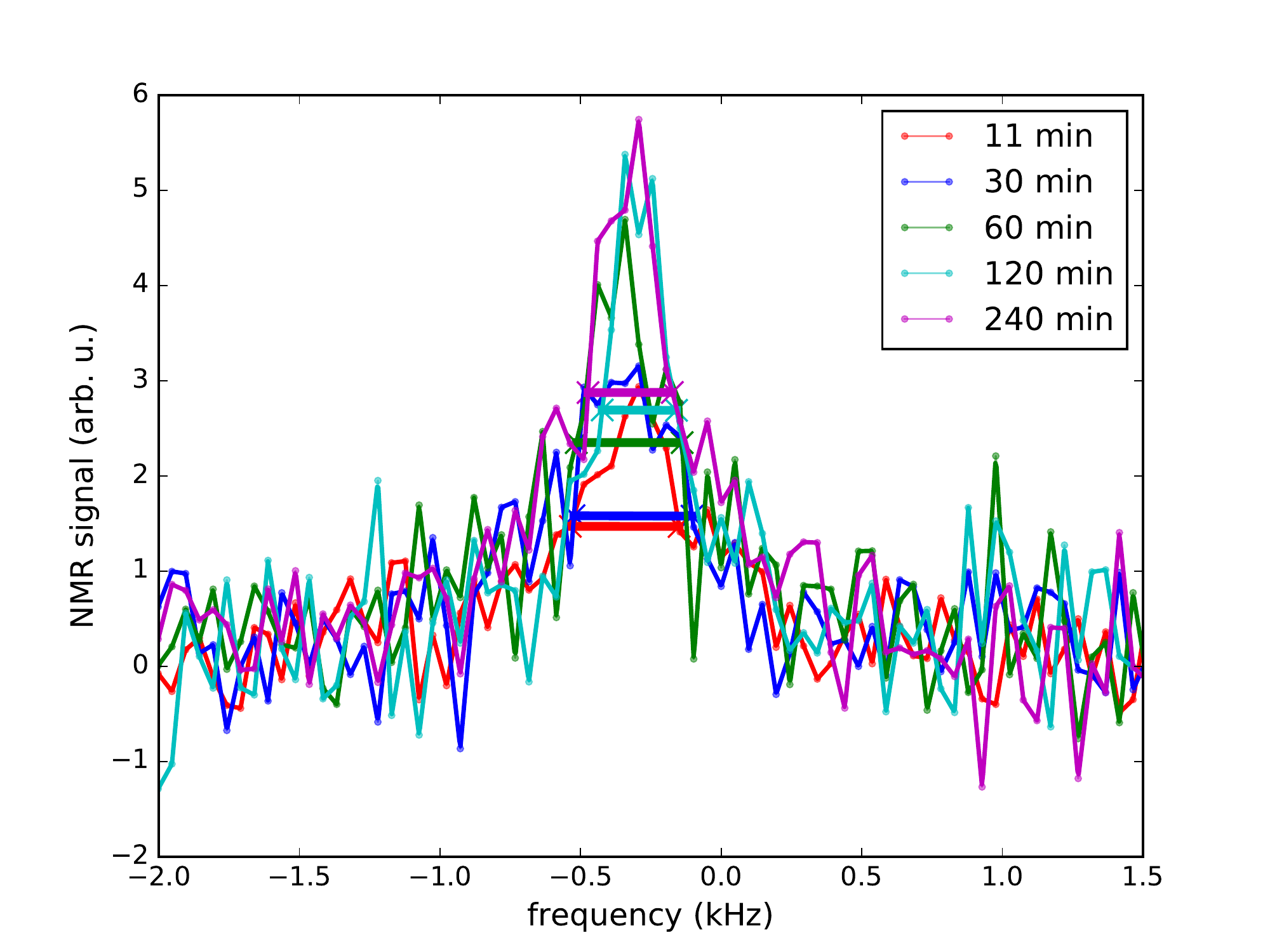}
\includegraphics[width=0.49\linewidth]{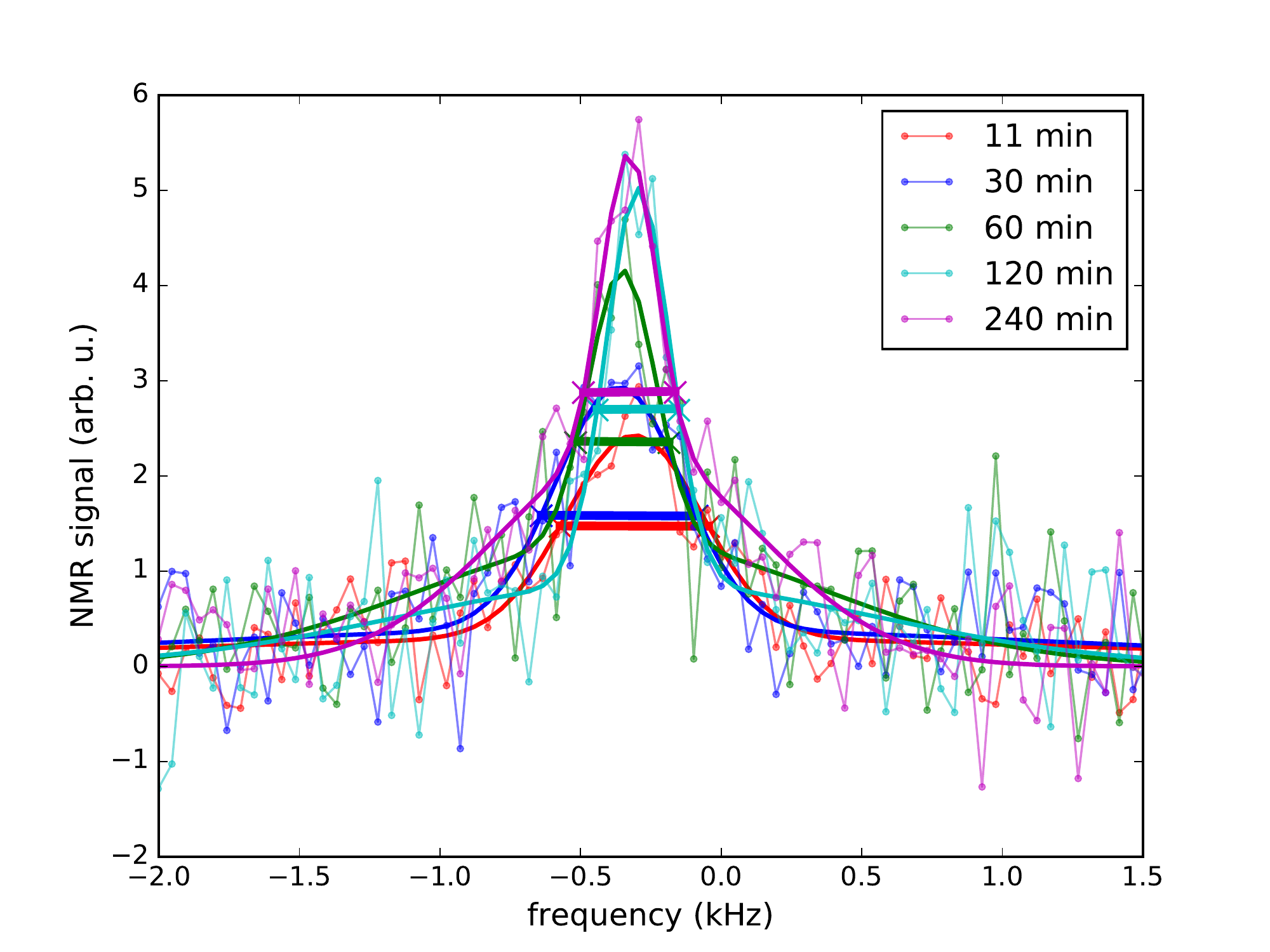}
\put(-465,155){(a)}
\put(-210,155){(b)}
\caption{Fourier transformed NMR signal (real part) for different last delay times. Extraction of the linewidth (FWHM) using a linear interpolation (a) as well as a double Gaussian (b).}
\label{300erSpectraFWHM}
\end{figure}

\begin{figure}
\tikzsetnextfilename{500erThermalSpectra}
\begin{tikzpicture}
	\begin{axis}[
	height=6cm, width=1.0\linewidth, xmin=0, xmax= 9, xlabel= frequency (\si{\kilo\hertz}), ylabel={NMR signal (arb. u.)}, legend style={at={(0.8,0.8)}, anchor=north,legend columns=2}, ylabel near ticks, yticklabel pos=left,]
\addplot[brown, no markers, line width= 1.0] table {data/export500er10sFit.dat};
\addlegendentry{\SI{10}{\second}};
\addplot[gray, no markers, line width= 1.0] table {data/export500er30sFit.dat};
\addlegendentry{\SI{30}{\second}};
\addplot[red, no markers, line width= 1.0] table {data/export500er100sFit.dat};
\addlegendentry{\SI{100}{\second}};
\addplot[blue, no markers, line width= 1.0] table {data/export500er300sFit.dat};
\addlegendentry{\SI{300}{\second}};
\addplot[green, no markers, line width= 1.0] table {data/export500er1000sFit.dat};
\addlegendentry{\SI{1000}{\second}};
\addplot[cyan, no markers, line width= 1.0] table {data/export500er3000sFit.dat};
\addlegendentry{\SI{3000}{\second}};
\addplot[magenta, no markers, line width= 1.0] table {data/export500er10000sFit.dat};
\addlegendentry{\SI{10000}{\second}};
\addplot[yellow, no markers, line width= 1.0] table {data/export500er30000sFit.dat};
\addlegendentry{\SI{30000}{\second}};
\addplot[ black, no markers, line width= 1.0] table {data/export500er260000sFit.dat};
\addlegendentry{\SI{260000}{\second}};
\addplot[brown, only marks, opacity=0.8, mark= o] table {data/export500er10s.dat};
\addplot[gray, only marks, opacity=0.8, mark= o] table {data/export500er30s.dat};
\addplot[red, only marks, opacity=0.8, mark= o] table {data/export500er100s.dat};
\addplot[blue, only marks, opacity=0.8, mark= o] table {data/export500er300s.dat};
\addplot[green, only marks, opacity=0.8, mark= o] table {data/export500er1000s.dat};
\addplot[cyan, only marks, opacity=0.8, mark= o] table {data/export500er3000s.dat};
\addplot[magenta, only marks, opacity=0.8, mark= o] table {data/export500er10000s.dat};
\addplot[yellow, only marks, opacity=0.8, mark= o] table {data/export500er30000s.dat};
\addplot[black, only marks, opacity=0.8, mark= o] table {data/export500er260000s.dat};
\end{axis}
\end{tikzpicture}
\caption{Fourier transformed NMR signal (real part) for different delay times after saturation. The solid lines correspond to a fitting to a Lorentzian function.}
\label{500erSpectra}
\end{figure}
Figure~\ref{500erSpectra} shows the NMR-signal in the frequency domain for different waiting times after the saturation pulses. The peaks are fitted to an Lorentzian function.
For the first two data points (\SI{10}{\second}, \SI{30}{\second}) \num{512} scans were accumulated, four scans for the data points from \SIrange{e2}{3e4}{\second} and a single measurement for \SI{26e4}{\second} ($\approx \SI{3}{\day}$).
\FloatBarrier
\section{Spin diffusion}
For the estimation of the diffusion range, we assume a radial isotropic spin diffusion around one hyperpolarized $^{13}$C spin \cite{Harris}.
\begin{align}
\begin{split}
\pdv{P(\vec{x},t)}{t} &= D \laplacian{P(\vec{x},t)} + (P_0 -P(\vec{x},t))R,~\text{mit~} P(\vec{x},t) \rightarrow P\mathrel{\mathop:}=P(r,\theta,\phi,t)\\
&= D \left(  \frac{1}{r^2} \pdv{}{r} \left( r^2 \pdv{P}{r} \right) + \frac{1}{r^2 \sin \theta} \pdv{}{\theta}\left(\sin \theta \pdv{P}{\theta} + \frac{1}{r^2 \sin^2 \theta}\pdv[2]{P}{\phi} \right) \right) \\
&+ (P_0 -P)R\\
&= D   \frac{1}{r^2} \pdv{}{r} \left( r^2 \pdv{P}{r} \right) + (P_0 -P)R
\end{split}
\label{eqDiffpolarisationRadial}
\end{align}
Here, $P_0$ (here $P_0= \num{e-8}$) is the thermal equilibrium polarization and $R=1/T_1$ the relaxation rate in the thermal equilibrium. The isotropic diffusion constant is $D=\SI{6,7e-15}{\square\centi\meter\per\second}$ \cite{Terblanche2001}.
In short:
\begin{equation}
\pdv{P(r,t)}{t} = D \frac{1}{r^2} \pdv{}{r} \left( r^2 \pdv{P(r,t)}{r} \right) + \left(P_0 -P(r,t)\right)R
\label{spindiffequation}
\end{equation}
For the initially polarized $^{13}$C spin a time behavior of $P(t,r_{\text{min}})=\exp{-0,5 t}$ is assumed.
This leads to a very short diffusion range of only several angstrom (Fig.~\ref{spindiff3D}).
In the supplemental material of Ref. \citenum{Fischer2013} a maximal diffusion length of $\SI{2.5}{\nano\meter}$ is estimated.
\begin{figure}[h]
\centering
\includegraphics[width=0.75\linewidth]{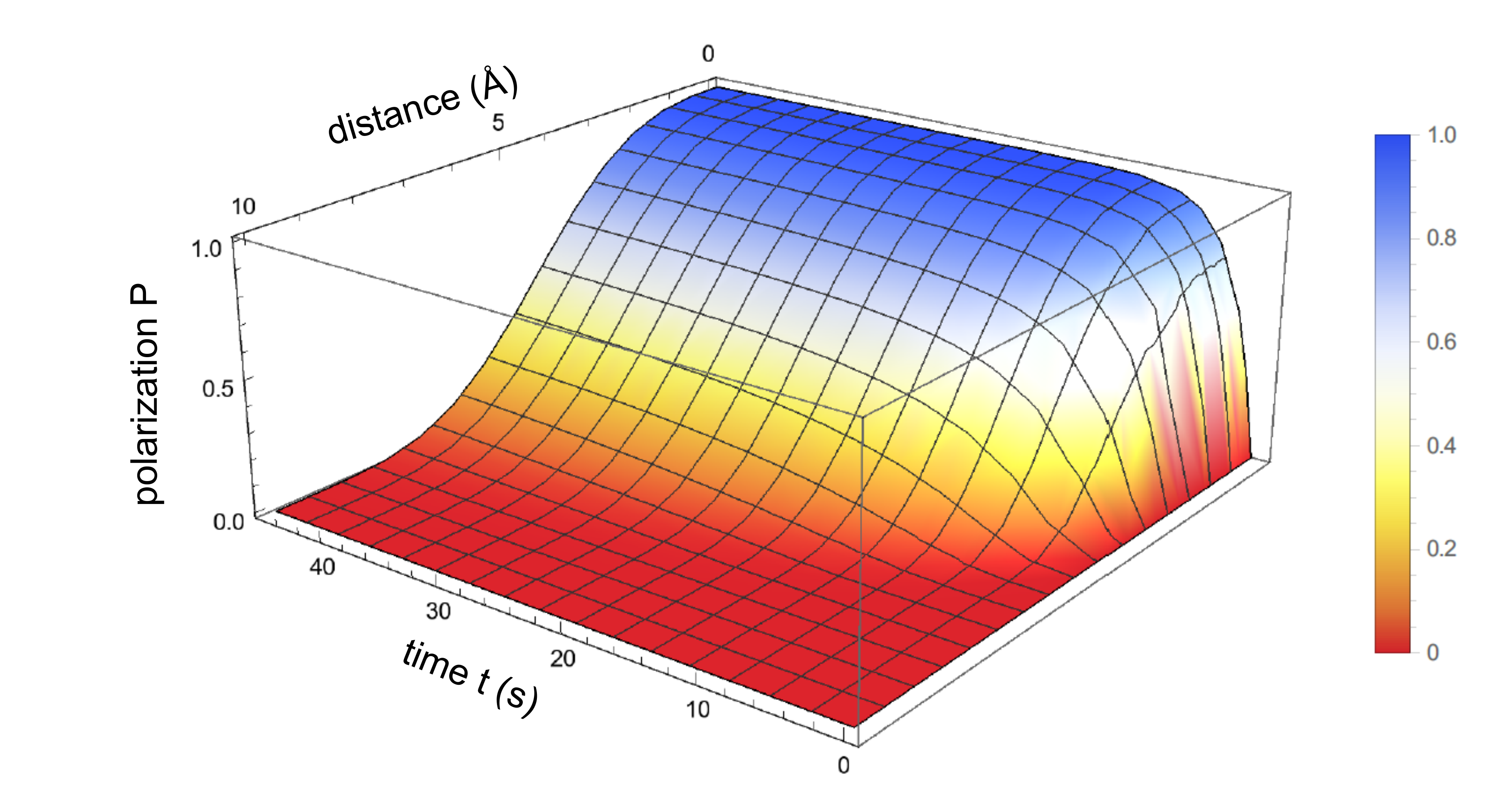}
\caption{3D plot of the solution of the spin diffusion equation \ref{spindiffequation}.}
\label{spindiff3D}
\end{figure}